%% file: main.tex
\errorstopmode
\input amssym.def
\input amssym.tex
\input format
\input epsf
\epsfclipon
\input macros

\input title

\input sect1

\input sect2

\input sect3

\input sect4

\input appa

\input appb

\input biblio
\bye

%% file: format

\magnification=\magstephalf
\hsize=14.0 true cm
\vsize=19 true cm
\hoffset=1.0 true cm
\voffset=2.0 true cm

\abovedisplayskip=12pt plus 3pt minus 3pt
\belowdisplayskip=12pt plus 3pt minus 3pt
\parindent=1.0em


\font\sixrm=cmr6
\font\eightrm=cmr8
\font\ninerm=cmr9

\font\sixi=cmmi6
\font\eighti=cmmi8
\font\ninei=cmmi9

\font\sixsy=cmsy6
\font\eightsy=cmsy8
\font\ninesy=cmsy9

\font\sixbf=cmbx6
\font\eightbf=cmbx8
\font\ninebf=cmbx9

\font\eightit=cmti8
\font\nineit=cmti9

\font\eightsl=cmsl8
\font\ninesl=cmsl9

\font\sixss=cmss8 at 8 true pt
\font\sevenss=cmss9 at 9 true pt
\font\eightss=cmss8
\font\niness=cmss9
\font\tenss=cmss10

\font\eighttt=cmtt8
\font\ninett=cmtt9
\font\tentt=cmtt10

\font\fivemib=cmmib5
\font\sixmib=cmmib6
\font\sevenmib=cmmib7
\font\eightmib=cmmib8
\font\ninemib=cmmib9
\font\tenmib=cmmib10

\font\fivesyb=cmbsy5
\font\sevensyb=cmbsy7
\font\tensyb=cmbsy10

 at 12 true pt
 at 12 true pt
\font\bigrm=cmr10 at 12 true pt
 at 12 true pt
 at 12 true pt

 at 16 true pt
 at 16 true pt
 at 16 true pt
 at 16 true pt
 at 16 true pt

\catcode`@=11
\newfam\mibfam
\newfam\ssfam

\def\tenpoint{\def\rm{\fam0\tenrm}%
    \textfont0=\tenrm \scriptfont0=\sevenrm \scriptscriptfont0=\fiverm
    \textfont1=\teni  \scriptfont1=\seveni  \scriptscriptfont1=\fivei
    \textfont2=\tensy \scriptfont2=\sevensy \scriptscriptfont2=\fivesy
    \textfont3=\tenex \scriptfont3=\tenex   \scriptscriptfont3=\tenex
    \textfont\itfam=\tenit                  \def\it{\fam\itfam\tenit}%
    \textfont\slfam=\tensl                  \def\sl{\fam\slfam\tensl}%
    \textfont\bffam=\tenbf \scriptfont\bffam=\sevenbf
                           \scriptscriptfont\bffam=\fivebf
                           \def\bf{\fam\bffam\tenbf}%
    \textfont\ssfam=\tenss \scriptfont\ssfam=\sevenss
                           \scriptscriptfont\ssfam=\sevenss
                           \def\ss{\fam\ssfam\tenss}%
    \textfont\ttfam=\tentt \def\tt{\fam\ttfam\tentt}%
    \textfont\mibfam=\tenmib \scriptfont\mibfam=\sevenmib
                             \scriptscriptfont\mibfam=\sevenmib
                             \def\mib{\fam\mibfam\tenmib}%
    \normalbaselineskip=13pt
    \setbox\strutbox=\hbox{\vrule height8.5pt depth3.5pt width0pt}%
    \let\big=\tenbig
    \normalbaselines\rm}

\def\ninepoint{\def\rm{\fam0\ninerm}%
    \textfont0=\ninerm      \scriptfont0=\sixrm
                            \scriptscriptfont0=\fiverm
    \textfont1=\ninei       \scriptfont1=\sixi
                            \scriptscriptfont1=\fivei
    \textfont2=\ninesy      \scriptfont2=\sixsy
                            \scriptscriptfont2=\fivesy
    \textfont3=\tenex       \scriptfont3=\tenex
                            \scriptscriptfont3=\tenex
    \textfont\itfam=\nineit \def\it{\fam\itfam\nineit}%
    \textfont\slfam=\ninesl \def\sl{\fam\slfam\ninesl}%
    \textfont\bffam=\ninebf \scriptfont\bffam=\sixbf
                            \scriptscriptfont\bffam=\fivebf
                            \def\bf{\fam\bffam\ninebf}%
    \textfont\ssfam=\niness \scriptfont\ssfam=\sixss
                            \scriptscriptfont\ssfam=\sixss
                            \def\ss{\fam\ssfam\niness}%
    \textfont\ttfam=\ninett \def\tt{\fam\ttfam\ninett}%
    \textfont\mibfam=\ninemib \scriptfont\mibfam=\sixmib
                            \scriptscriptfont\mibfam=\sixmib
                            \def\mib{\fam\mibfam\ninemib}%
    \normalbaselineskip=12pt
    \setbox\strutbox=\hbox{\vrule height8.0pt depth3.0pt width0pt}%
    \let\big=\ninebig
    \normalbaselines\rm}

\def\eightpoint{\def\rm{\fam0\eightrm}%
    \textfont0=\eightrm      \scriptfont0=\sixrm
                             \scriptscriptfont0=\fiverm
    \textfont1=\eighti       \scriptfont1=\sixi
                             \scriptscriptfont1=\fivei
    \textfont2=\eightsy      \scriptfont2=\sixsy
                             \scriptscriptfont2=\fivesy
    \textfont3=\tenex        \scriptfont3=\tenex
                             \scriptscriptfont3=\tenex
    \textfont\itfam=\eightit \def\it{\fam\itfam\eightit}%
    \textfont\slfam=\eightsl \def\sl{\fam\slfam\eightsl}%
    \textfont\bffam=\eightbf \scriptfont\bffam=\sixbf
                             \scriptscriptfont\bffam=\fivebf
                             \def\bf{\fam\bffam\eightbf}%
    \textfont\ssfam=\eightss \scriptfont\ssfam=\sixss
                             \scriptscriptfont\ssfam=\sixss
                             \def\ss{\fam\ssfam\eightss}%
    \textfont\ttfam=\eighttt \def\tt{\fam\ttfam\eighttt}%
    \textfont\mibfam=\eightmib \scriptfont\mibfam=\sixmib
                             \scriptscriptfont\mibfam=\sixmib
                             \def\mib{\fam\mibfam\eightmib}%
    \normalbaselineskip=10pt
    \setbox\strutbox=\hbox{\vrule height7.0pt depth2.0pt width0pt}%
    \let\big=\eightbig
    \normalbaselines\rm}

\def\tenbig#1{{\hbox{$\left#1\vbox to8.5pt{}\right.\n@space$}}}
\def\ninebig#1{{\hbox{$\textfont0=\tenrm\textfont2=\tensy
                       \left#1\vbox to7.25pt{}\right.\n@space$}}}
\def\eightbig#1{{\hbox{$\textfont0=\ninerm\textfont2=\ninesy
                       \left#1\vbox to6.5pt{}\right.\n@space$}}}

\def\setmathbold{
\textfont1=\tenmib \scriptfont1=\sevenmib \scriptscriptfont1=\fivemib
\textfont2=\tensyb \scriptfont2=\sevensyb \scriptscriptfont2=\fivesyb}

\def\unsetmathbold{
\textfont1=\teni \scriptfont1=\seveni \scriptscriptfont1=\fivei
\textfont2=\tensy \scriptfont2=\sevensy \scriptscriptfont2=\fivesy
}

\font\sectionfont=cmbx10
\font\subsectionfont=cmti10

\def\figurecaptionfont{\ninepoint}
\def\tablecaptionfont{\ninepoint}
\def\footnotefont{\eightpoint}


\newcount\equationno
\newcount\bibitemno
\newcount\figureno
\newcount\tableno

\equationno=0
\bibitemno=0
\figureno=0
\tableno=0


\footline={\ifnum\pageno=0{\hfil}\else
{\hss\rm\the\pageno\hss}\fi}


\def\section #1. #2 \par
{\vskip0pt plus .10\vsize\penalty-100 \vskip0pt plus-.10\vsize
\vskip 1.6 true cm plus 0.2 true cm minus 0.2 true cm
\global\def\equationlabel{#1}
\global\equationno=0
\setmathbold
\leftline{\sectionfont #1. #2}\par
\immediate\write\terminal{Section #1. #2}\unsetmathbold
\vskip 0.7 true cm plus 0.1 true cm minus 0.1 true cm
\noindent}


\def\subsection #1 \par
{\vskip0pt plus 1.0 true cm\penalty-50 \vskip0pt plus-1.0 true cm
\vskip2.5ex plus 0.1ex minus 0.1ex
\leftline{\subsectionfont #1}\par
\immediate\write\terminal{Subsection #1}
\vskip1.0ex plus 0.1ex minus 0.1ex
\noindent}


\def\appendix #1. #2 \par
{\vskip0pt plus .10\vsize\penalty-100 \vskip0pt plus-.10\vsize
\vskip 1.6 true cm plus 0.2 true cm minus 0.2 true cm
\global\def\equationlabel{\hbox{\rm#1}}
\global\equationno=0
\setmathbold
\leftline{\sectionfont Appendix #1. #2}\par
\immediate\write\terminal{Appendix #1. #2}\unsetmathbold
\vskip 0.7 true cm plus 0.1 true cm minus 0.1 true cm
\noindent}



\def\equation#1{$$\displaylines{\qquad #1}$$}
\def\enum{\global\advance\equationno by 1
\hfill\llap{{\rm(\equationlabel.\the\equationno)}}}

\def\nexteq#1{\cr\noalign{\vskip#1}\qquad}


\def\ifundefined#1{\expandafter\ifx\csname#1\endcsname\relax}

\def\ref#1{\ifundefined{#1}?\immediate\write\terminal{unknown reference
on page \the\pageno}\else\csname#1\endcsname\fi}

\newwrite\terminal
\newwrite\bibitemlist

\def\bibitem#1#2\par{\global\advance\bibitemno by 1
\immediate\write\bibitemlist{\string\def
\expandafter\string\csname#1\endcsname
{\the\bibitemno}}
\item{[\the\bibitemno]}#2\par}

\def\beginbibliography{
\vskip0pt plus .15\vsize\penalty-100 \vskip0pt plus-.15\vsize
\vskip 1.2 true cm plus 0.2 true cm minus 0.2 true cm
\leftline{\sectionfont References}\par
\immediate\write\terminal{References}
\immediate\openout\bibitemlist=biblist
\frenchspacing\parindent=1.8em
\vskip 0.5 true cm plus 0.1 true cm minus 0.1 true cm}

\def\endbibliography{
\immediate\closeout\bibitemlist
\nonfrenchspacing\parindent=1.0em}


\def\figurecaption#1{\global\advance\figureno by 1
\narrower\figurecaptionfont Fig.~\the\figureno. #1}

\def\tablecaption#1{\global\advance\tableno by 1
\centerline{\tablecaptionfont Table~\the\tableno. #1}}

\def\thicktablerule{\hrule height0.8pt}
\def\thintablerule{\hrule height0.4pt}

\tenpoint

\immediate\openin\bibitemlist=biblist
\ifeof\bibitemlist\immediate\closein\bibitemlist
\else\immediate\closein\bibitemlist

\input biblist \fi


\def\thismonth{\ifcase\month\or
January\or February\or March\or April\or May\or June\or
July\or August\or September\or October\or November\or December\fi}

\font\fourteenrm=cmr12 scaled \magstep1
\font\fourteeni=cmmi12 scaled \magstep1
\font\fourteensy=cmsy10 scaled \magstep2
\font\fourteenex=cmex10 scaled \magstep2
\font\fourteenbf=cmbx12 scaled \magstep1
\font\fourteensl=cmsl12 scaled \magstep1
\font\fourteentt=cmtt12 scaled \magstep1
\font\fourteenit=cmti12 scaled \magstep1

\ifx\fourteenpoint\undefined
   \def\fourteenpoint{\def\rm{\fam0\fourteenrm}
       \textfont0=\fourteenrm \scriptfont0=\tenrm \scriptscriptfont0=\sevenrm
       \textfont1=\fourteeni  \scriptfont1=\teni  \scriptscriptfont1=\seveni
       \textfont2=\fourteensy \scriptfont2=\tensy \scriptscriptfont2=\sevensy
       \textfont3=\fourteenex \scriptfont3=\fourteenex
                              \scriptscriptfont3=\fourteenex
       \textfont\itfam=\fourteenit  \def\it{\fam\itfam\fourteenit}%
       \textfont\slfam=\fourteensl  \def\sl{\fam\slfam\fourteensl}%
       \textfont\ttfam=\fourteentt  \def\tt{\fam\ttfam\fourteentt}%
       \textfont\bffam=\fourteenbf  \scriptfont\bffam=\tenbf
        \scriptscriptfont\bffam=\sevenbf  \def\bf{\fam\bffam\fourteenbf}%
       \normalbaselineskip=17pt
       \setbox\strutbox=\hbox{\vrule height11.9pt depth6.3pt width0pt}%
       \normalbaselines\rm}
   \fi

\font\seventeenrm=cmr17
\font\seventeeni=cmmi12 scaled \magstep2
\font\seventeensy=cmsy10 scaled \magstep3
\font\seventeenex=cmex10 scaled \magstep3
\font\seventeenbf=cmbx12 scaled \magstep2
\font\seventeensl=cmsl12 scaled \magstep2
\font\seventeentt=cmtt12 scaled \magstep2
\font\seventeenit=cmti12 scaled \magstep2

\font\twelverm=cmr12
\font\twelvei=cmmi12
\font\twelvesy=cmsy10 scaled \magstep1
\font\twelvebf=cmbx12

\font\ninerm=cmr9
\font\ninei=cmmi9
\font\ninesy=cmsy9
\font\ninebf=cmbx9

\ifx\seventeenpoint\undefined
   \def\seventeenpoint{\def\rm{\fam0\seventeenrm}
       \textfont0=\seventeenrm \scriptfont0=\twelverm \scriptscriptfont0=\ninerm
       \textfont1=\seventeeni  \scriptfont1=\twelvei  \scriptscriptfont1=\ninei
       \textfont2=\seventeensy \scriptfont2=\twelvesy \scriptscriptfont2=\ninesy
       \textfont3=\seventeenex \scriptfont3=\seventeenex
                              \scriptscriptfont3=\seventeenex
       \textfont\itfam=\seventeenit  \def\it{\fam\itfam\seventeenit}%
       \textfont\slfam=\seventeensl  \def\sl{\fam\slfam\seventeensl}%
       \textfont\ttfam=\seventeentt  \def\tt{\fam\ttfam\seventeentt}%
       \textfont\bffam=\seventeenbf  \scriptfont\bffam=\twelvebf
        \scriptscriptfont\bffam=\ninebf  \def\bf{\fam\bffam\seventeenbf}%
       \normalbaselineskip=21pt
       \setbox\strutbox=\hbox{\vrule height17pt depth4pt width0pt}%
       \normalbaselines\rm}
   \fi

%% file: macros


\def\rmd{{\rm d}}

\def\rme{{\rm e}}
\def\rmO{{\rm O}}
\def\rmU{{\rm U}}

\def\urltilde{\kern -.15em\lower .7ex\hbox{\~{}}\kern .04em}


\def\gz{{\Bbb Z}}

\def\Re{{\rm Re}\,}


\def\proof{\noindent{\sl Proof:}\kern0.6em}

\def\frac#1#2{\hbox{$#1\over#2$}}
\def\dual{\mathstrut^*\kern-0.1em}

\def\lvec#1{\setbox0=\hbox{$#1$}
    \setbox1=\hbox{$\scriptstyle\leftarrow$}
    #1\kern-\wd0\smash{
    \raise\ht0\hbox{$\raise1pt\hbox{$\scriptstyle\leftarrow$}$}}
    \kern-\wd1\kern\wd0}
\def\rvec#1{\setbox0=\hbox{$#1$}
    \setbox1=\hbox{$\scriptstyle\rightarrow$}
    #1\kern-\wd0\smash{
    \raise\ht0\hbox{$\raise1pt\hbox{$\scriptstyle\rightarrow$}$}}
    \kern-\wd1\kern\wd0}
\def\cvec#1{\kern-0.5pt\vec{\kern0.5pt #1}}

\def\slash#1{\setbox2=\hbox{$\displaystyle#1$}%
             \setbox3=\hbox{$\displaystyle/$}%
             #1\kern-0.8\wd2/\kern-1.0\wd3\kern0.8\wd2\kern0.5pt}

\def\wick#1{\setbox2=\hbox{$\displaystyle#1$}
    \setbox3=\null\ht3=3.0pt\dp3=0.0pt\wd3=20.0pt
    #1\kern-\wd2\kern3.0pt\raise11.0pt\vbox{\hrule height0.3pt
    \hbox{\vrule width0.3pt\box3\vrule width0.3pt}}\kern-24.0pt\kern\wd2}

\def\longwick#1{\setbox2=\hbox{$\displaystyle#1$}
    \setbox3=\null\ht3=3.0pt\dp3=0.0pt\wd3=27.0pt
    #1\kern-\wd2\kern3.0pt\raise11.0pt\vbox{\hrule height0.3pt
    \hbox{\vrule width0.3pt\box3\vrule width0.3pt}}\kern-31.0pt\kern\wd2}

\def\verylongwick#1{\setbox2=\hbox{$\displaystyle#1$}
    \setbox3=\null\ht3=3.0pt\dp3=0.0pt\wd3=43.0pt
    #1\kern-\wd2\kern3.0pt\raise11.0pt\vbox{\hrule height0.3pt
    \hbox{\vrule width0.3pt\box3\vrule width0.3pt}}\kern-47.0pt\kern\wd2}


\def\nabstar#1{{\nabla\kern0.5pt\smash{\raise 4.5pt\hbox{$\ast$}}
               \kern-5.5pt_{#1}}}

\def\drvstar#1{{\partial\kern0.5pt\smash{\raise 4.5pt\hbox{$\ast$}}
               \kern-6.0pt_{#1}}}
\def\sdrvstar#1{{\partial\kern0.4pt\smash{\raise 3.6pt\hbox{$\ast$}}
                \kern-4.8pt_{#1}}}

\def\ldrvstar#1{{\lvec{\,\partial}\kern-0.5pt\smash{\raise 4.5pt\hbox{$\ast$}}
               \kern-5.0pt_{#1}}}


\def\MeV{{\rm MeV}}

\def\fm{{\rm fm}}
\def\MSbar{\overline{\rm MS\kern-0.5pt}\kern0.5pt}




\def\diracstar#1#2{
    \setbox0=\hbox{$\gamma$}\setbox1=\hbox{$\gamma_{#1}$}
    \gamma_{#1}\kern-\wd1\kern\wd0
    \smash{\raise4.5pt\hbox{$\scriptstyle#2$}}}


\def\SUthree{{\rm SU(3)}}

\def\tr{{\rm tr}}



\def\Tc{T_{\rm c}}

\def\xv{{\mib x}}

%% file: title
%
\rightline{CERN-TH-2018-259}
\vskip1.2cm
{\fourteenpoint
\centerline{Topological susceptibility at $T>\Tc$ from master-field}
\vskip 0.2 true cm
\centerline{simulations of the SU(3) gauge theory}
}

\tenpoint
\vskip 0.6 true cm
\centerline{\bigrm Leonardo Giusti$^{\rm a}$ and
Martin L\"uscher$^{\rm b,c}$}

\vskip1.5ex
\centerline{{\it $^{\rm a}$Dipartimento di Fisica,
Universit\`a di Milano-Bicocca and}}
\centerline{{\it INFN, Sezione di Milano-Bicocca,
Piazza della Scienza 3, I-20126 Milano, Italy}}

\vskip1.0ex
\centerline{{\it $^{\rm b}$CERN,
Theoretical Physics Department, 1211 Geneva 23, Switzerland}}

\vskip1.0ex
\centerline{{\it $^{\rm c}$Albert Einstein Center for Fundamental Physics}}
\centerline{{\it Institute for Theoretical Physics,
Sidlerstrasse 5, 3012 Bern, Switzerland}}

\vskip 0.8 true cm
\thintablerule
\vskip 2.0ex
\ninepoint
\leftline{\bf Abstract}
\vskip 1.0ex\noindent
The topological susceptibility is
computed in the $\SUthree$ gauge theory at temperatures $T$
above the critical temperature $\Tc$
using master-field simulations of very large lattices, where
the infamous topology-freezing issue is effectively bypassed.
Up to $T=2.0\,\Tc$ no unusually large lattice effects are observed and
the results obtained in the continuum limit confirm the
expected rapid decay of the susceptibility with increasing temperature.
As a byproduct, the reference gradient-flow time $t_0$ is
determined in the range of lattice spacings from
$0.023$ to $0.1\,\fm$ with a precision of 2 per mille.
\vskip 2.0ex
\thintablerule

\tenpoint


%% file: sect1
\section 1. Introduction

The temperature dependence of the topological susceptibility $\chi_t$ in QCD
is of interest in connection with the dark-matter candidacy
of the axion, a hypothetical particle related to the so-called strong
CP problem [\ref{PecceiQuinnI}--\ref{Wilczek}].
Computations of $\chi_t$
in numerical lattice QCD are however not
straightforward for various reasons.
A direct sampling of the topological charge
is often impractical, for example,
because the simulation algorithms tend to get
trapped in a fixed-charge sector of field space.
Another source of difficulty is the fact that
the susceptibility decreases rapidly at high temperatures
and consequently becomes more and more sensitive to lattice effects.

Most computations of the topological susceptibility at temperatures $T$
larger than the critical temperature $\Tc$ performed
to date [\ref{BerkowitzEtAl}--\ref{BonatiEtAlII}]
rely on some form of reweighting or the so-called integral
method, where $\chi_t$ is obtained by integrating its derivative
with respect to $T$ from low to high temperatures.
The systematic uncertainties and the statistical errors
are generally fairly large in these calculations,
particularly so when the light quarks
(which lead to an additional chiral suppression of $\chi_t$) are included.

Master-field simulations [\ref{LuscherMaster}]
bypass the topology freezing issue by simulating lattices with
four-dimensional volumes $V$ satisfying
\equation{
  \chi_tV\gg 1.
  \enum
}
Fixed-topology effects are of order $1/V$ in this case
[\ref{BrowerEtAl},\ref{AokiEtAl}] and are
thus parametrically smaller than the statistical errors, which
decrease like $V^{-1/2}$ at large $V$.
In the present paper, master-field simulations are used
to calculate the topological susceptibility in the SU(3) gauge
theory at temperatures approximately equal to $1.5\,\Tc$ and
$2.0\,\Tc$. The study also serves as a first test of the feasibility
of such simulations at non-zero temperatures, where having
a physically large three-dimensional volume may be
of some general interest.

In the next section, the theoretical framework is described in more detail.
Since the topological susceptibility is rapidly varying with
temperature, its extrapolation to the continuum limit requires
a highly accurate scale setting. A separate computation of the
reference gradient-flow time $t_0$ [\ref{WilsonFlow}]
was therefore performed using master-field simulations
at vanishing temperature. The computation of $\chi_t$
is discussed in sect.~3 and conclusions are drawn in sect.~4.

%% file: sect2
\section 2. Theoretical framework

\vskip-2.5ex

\subsection 2.1 Lattice theory

The SU(3) Yang--Mills theory studied in this paper is set up
on hyper-cubic $L_0\times L^3$ lattices
with spacing $a$ and periodic boundary conditions in all directions.
At high temperatures $T=1/L_0$, the time extent $L_0$ of the lattice
is always taken to be much smaller than its spatial size $L$.
For the gauge action the Wilson plaquette action [\ref{Wilson}] with bare
coupling $g_0$ is chosen.

\subsection 2.2 Definition of $\chi_t$

Since the correlation function of the topological density
\equation{
  q(x)=-{1\over32\pi^2}\,\epsilon_{\mu\nu\rho\sigma}
  \tr\{F_{\mu\nu}(x)F_{\rho\sigma}(x)\}
  \enum
}
(where $F_{\mu\nu}$ denotes the field strength of the
gauge potential) has a non-integrable short-distance singularity,
the topological susceptibility is only formally given by
\equation{
  \chi_t=\int\rmd^4 x\,\langle q(x)q(0)\rangle.
  \enum
}
A sensible definition of the susceptibility in the continuum
theory must therefore be provided
before it can be computed on the lattice.

In the present context, the susceptibility is tied to
the flavour-singlet $\rmU(1)$ chiral symmetry of QCD,
which becomes a non-anomalous symmetry
when the axion field is included in the theory.
The soft breaking of the symmetry by the quark masses
then leads to the well-known formula relating the
axion mass to $\chi_t$, provided the latter
is defined consistently with the chiral Ward identities.
When this condition is met, $\chi_t$ is unambiguously
determined and can be shown to be given
by a singularity-free expectation value of ``density chains''
[\ref{GiustiEtAlI}--\ref{LuscherShort}].

Far easier to evaluate than the density chains is the topological
charge at positive gradient-flow time [\ref{WilsonFlow}].
The associated susceptibility does not require any subtraction
or renormalization [\ref{RenFlow}] and is known to coincide
with the susceptibility defined through the density chains,
at least in the pure gauge theory [\ref{CeEtAl}].
All this holds in the continuum limit of the lattice theory,
provided the flow time is held fixed in physical units when
the lattice spacing is taken to zero.
In the present paper, the topological susceptibility is measured
in this way, the implementation of the gradient flow and
other technical details being the same as in ref.~[\ref{WilsonFlow}].

\subsection 2.3 Physical regimes at high temperatures

The topological susceptibility is a potentially complicated
function of the temperature $T$ and the spatial volume $L^3$,
particularly so when $L$ is less than $1\,\fm$, where the
effective gauge coupling is small and the semi-classical
approximation becomes asymptotically
exact\kern1pt\footnote{$\dagger$}{\footnotefont%
In the case of a four-dimensional spherical space-time,
$\chi_t$ can be worked out analytically in this limit and is found to
be a steep function of $V$ [\ref{SemiClassical}].
At non-zero temperatures,
the situation is far more complicated
already at the classical level
[\ref{BaalKraanI},\ref{BaalKraanII}].}.
If $L$ is much larger than the correlation lengths in
the pseudo-scalar sector, $\chi_t$ is independent of $L$
up to exponentially small terms. This regime
sets in at values of $L$ of a few fermi, for all temperatures,
but at high temperatures the bound (1.1)
only holds at much larger spatial sizes.

At these temperatures there is then an interesting
intermediate regime, in which
$L$ is large while the variance
\equation{
  \langle Q^2\rangle=\chi_tV,
  \qquad V=L^3/T,
  \enum
}
of the distribution of the topological charge $Q$ is much smaller
than $1$.
It is plausible that $\chi_t$ is dominated by the sectors with
charge $Q=\pm1$ in this case. Moreover, if
their contribution is assumed to be suppressed by the factor
$\exp\{-S_{\rm min}\}$, $S_{\rm min}$ being the minimum of the
gauge action in these sectors (the instanton action),
the renormalization group implies that
\equation{
  \chi_t\mathrel{\mathop\propto_{T\to\infty}}T^{-7}
  \enum
}
with a logarithmically varying proportionality constant.
It goes without saying that this
argumentation is quite crude and that eq.~(2.4) should not
be taken as a solid theoretical result.

\input table1

\subsection 2.4 Computation of the reference flow time $t_0$

The extrapolation to the continuum limit of lattice results
for the topological susceptibility requires a precise scale-setting.
When the limit is taken, the temperature must be held fixed in
units of some physical scale such as the Sommer radius
[\ref{SommerScale}]. Moreover, since $\chi_t$ has
mass dimension $4$, its value must also be expressed in such units.
In view of the steep temperature dependence of $\chi_t$, a relative
numerical error in the reference scale thus results in an approximately
$11$ times larger error of the converted values of $\chi_t$.

The target statistical precision of $\chi_t$ in the present paper
is a few percent and the reference scale must therefore be known with
errors less than a few per mille to permit unbiased continuum-limit
extrapolations. This level of precision
is generally difficult to reach in practice, but can be
attained with a limited computational effort
if the reference gradient-flow time $t_0$ [\ref{WilsonFlow}]
is used to set the scale.

\input figure1

The values of $t_0/a^2$ quoted in table~1 were obtained from
master-field simulations of physically large lattices.
In the range of $\beta=6/g_0^2$ considered,
the lattice spacing decreases from about $0.10$ to $0.023\,\fm$.
The lattice sizes $L$ are at least $6\,\fm$ and reach values
above $9\,\fm$ in some cases.
On all these lattices, $\chi_t V$ is in the thousands
and frozen-topology effects are therefore expected to be
neglible.
The numbers $N_{\rm mf}$ of master fields included in the
measurement of $t_0/a^2$
were adjusted so as to have approximately
constant statistical errors of about $2$ per mille.
Further details of the simulations are reported
in appendix A.

As shown in fig.~1, the data for $\ln(t_0/a^2)$ rise roughly linearly
with $\beta$ and can be well represented by a polynomial
\equation{
  \ln(t_0/a^2)=\sum_{k=0}^4c_k(\beta-6)^k
  \enum
}
of degree $4$.
A least-squares fit yields the values
\equation{
  (c_0,\ldots,c_4)=(1.16390,3.37888,-1.36231,1.20666,-0.45672)
  \enum
}
for the coefficients. The fit approximates $t_0/a^2$
in the range $5.96\leq\beta\leq7.01$ with an estimated error
of $2$ per mille.
A comparison with more precise results previously
obtained on small lattices [\ref{CeEtAl}] confirms this
up to $\beta=6.42$ and
the fit also reproduces the values at $\beta=6.3,6.4,\ldots,7.0$
quoted in ref.~[\ref{AsakawaEtAl}]
within errors varying from $0.2$ to $1.1$ percent.

\subsection 2.5 Conversion to physical units

The $\SUthree$ Yang--Mills theory is unphysical and any assignment of
physical units is therefore a bit arbitrary. Often the Sommer
radius $r_0$ is taken as the reference scale and its physical
value is set to $0.5\,\fm$. In the range $5.96\leq\beta\leq6.92$
of validity of the fit curves of both $r_0/a$ [\ref{NeccoSommer}]
and $t_0/a^2$, the ratio of scales plotted in fig.~1 averages to $0.950$.
The traditional choice $r_0=0.5\,\fm$ thus amounts to setting
\equation{
  (8t_0)^{1/2}=0.475\,\fm.
  \enum
}
Throughout this paper the conversion to physical units is
performed using eq.~(2.7) and the values of $t_0/a^2$ given
by the interpolation (2.5).

%% file: table1
\topinsert
\newdimen\digitwidth
\setbox0=\hbox{\rm 0}
\digitwidth=\wd0
\catcode`@=\active
\def@{\kern\digitwidth}
\tablecaption{Lattice parameters and simulation results for $t_0/a^2$}
\vskip-3.5ex

$$\vbox{\settabs\+&%
                  xxxxxxx&xx&%
                  xxxxx&xx&%
                  xxx&xx&%
                  xxxxxxxxxxx&xxxxx&%
                  xxxxxxx&xx&%
                  xxxxx&xx&%
                  xxx&xx&%
                  xxxxxxxxxx&x\cr
\thicktablerule
\vskip1.2ex
                \+& \hfill Lattice\hfill
                 && \hfill $\beta$\hfill
                 && \hfill $N_{\rm mf}$\hfill
                 && \hfill $t_0/a^2$\hfill
                 && \hfill Lattice\hfill
                 && \hfill $\beta$\hfill
                 && \hfill $N_{\rm mf}$\hfill
                 && \hfill $t_0/a^2$\hfill
                 &\cr
\vskip0.8ex
\thintablerule
\vskip1.2ex
{\ninepoint
  \+& \hfill $96^4$\hfill
  &&  \hfill $5.96$\hfill
  &&  \hfill $1$\hfill
  &&  \hfill $2.7875(53)$\hfill
  &&  \hfill $192^4$\hfill
  &&  \hfill $6.53$\hfill
  &&  \hfill $2$\hfill
  &&  \hfill $15.156(28)$\hfill
  &\cr
  \+& \hfill $96^4$\hfill
  &&  \hfill $6.05$\hfill
  &&  \hfill $4$\hfill
  &&  \hfill $3.7834(47)$\hfill
  &&  \hfill $192^4$\hfill
  &&  \hfill $6.61$\hfill
  &&  \hfill $4$\hfill
  &&  \hfill $18.714(30)$\hfill
  &\cr
  \+& \hfill $96^4$\hfill
  &&  \hfill $6.13$\hfill
  &&  \hfill $8$\hfill
  &&  \hfill $4.8641(85)$\hfill
  &&  \hfill $192^4$\hfill
  &&  \hfill $6.69$\hfill
  &&  \hfill $5$\hfill
  &&  \hfill $23.089(48)$\hfill
  &\cr
  \+& \hfill $96^4$\hfill
  &&  \hfill $6.21$\hfill
  &&  \hfill $4$\hfill
  &&  \hfill $6.219(13)@$\hfill
  &&  \hfill $192^4$\hfill
  &&  \hfill $6.77$\hfill
  &&  \hfill $6$\hfill
  &&  \hfill $28.494(66)$\hfill
  &\cr
  \+& \hfill $128^4$\hfill
  &&  \hfill $6.29$\hfill
  &&  \hfill $3$\hfill
  &&  \hfill $7.785(14)@$\hfill
  &&  \hfill $256^4$\hfill
  &&  \hfill $6.85$\hfill
  &&  \hfill $3$\hfill
  &&  \hfill $34.819(84)$\hfill
  &\cr
  \+& \hfill $128^4$\hfill
  &&  \hfill $6.37$\hfill
  &&  \hfill $5$\hfill
  &&  \hfill $9.755(19)@$\hfill
  &&  \hfill $256^4$\hfill
  &&  \hfill $6.93$\hfill
  &&  \hfill $5$\hfill
  &&  \hfill $42.82(11)@$\hfill
  &\cr
  \+& \hfill $128^4$\hfill
  &&  \hfill $6.42$\hfill
  &&  \hfill $7$\hfill
  &&  \hfill $11.202(21)@@$\hfill
  &&  \hfill $256^4$\hfill
  &&  \hfill $7.01$\hfill
  &&  \hfill $7$\hfill
  &&  \hfill $52.25(13)@$\hfill
  &\cr
  \+& \hfill $128^4$\hfill
  &&  \hfill $6.45$\hfill
  &&  \hfill $11$\hfill
  &&  \hfill $12.196(21)@@$\hfill
  &&  \hfill $$\hfill
  &&  \hfill $$\hfill
  &&  \hfill $$\hfill
  &&  \hfill $$\hfill
  &\cr
}
\vskip1.0ex
\thicktablerule
}
$$
\vskip-1.0ex
\endinsert

%% file: figure1
\topinsert
\vbox{
\vskip0ex

\epsfxsize=11.0cm
\hskip0.7cm\epsfbox{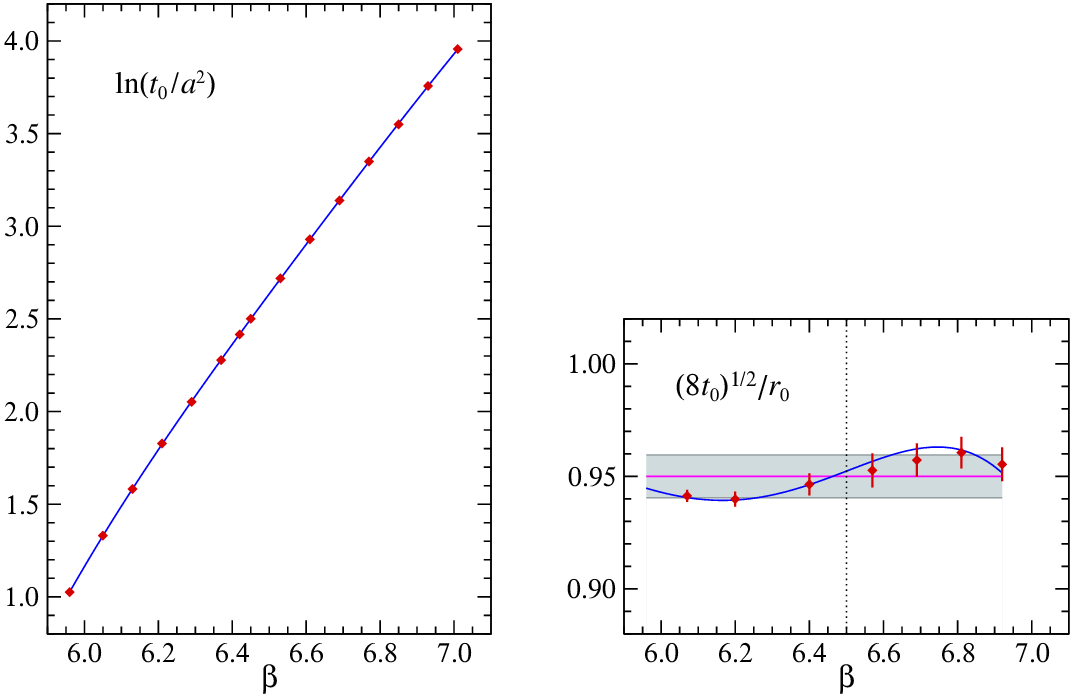}%

\vskip2.0ex
\figurecaption{%
Plot of the simulation results for $\ln(t_0/a^2)$ (diamonds) and
the interpolation (2.5),(2.6).
As shown by the plot on the right,
setting the scale with $t_0$ or the
available data for the Sommer radius $r_0$
[\ref{GuagnelliEtAl},\ref{NeccoSommer}] comes to the same
within a margin of about $1\%$ (grey band;
$r_0$ was computed using different methods above and below $\beta=6.5$).
The sinusoidal curve is obtained from the fit function
(2.5) and the one published by Necco and Sommer
for $r_0/a$ [\ref{NeccoSommer}].
}
}
\endinsert

%% file: sect3
\input table2

\section 3. Computation of the topological susceptibility

The computations reported in this section follow the lines of
refs.~[\ref{LuscherMaster},\ref{WilsonFlow}] except for the fact that
lattices at high temperatures are simulated.

\subsection 3.1 Master-field simulations

In total six lattices were simulated, at two temperatures and three
lattice spacings at each temperature, so as to allow for an
extrapolation of the results to the continuum limit
(see table~2). The critical temperature $\Tc$ in the SU(3) gauge
theory is $294\,\MeV$ [\ref{BoydEtAl}] and the chosen
temperatures $T$ are thus about $1.5\,\Tc$ and $2.0\,\Tc$.
As will become clear below,
the bound (1.1) is well satisfied on all lattices.
Moreover, the relevant
correlation lengths are much smaller than the spatial sizes $L$,
so that the master-field simulation strategy
is expected to work out.

\input figure2

At high temperatures, the Polyakov loop
\equation{
  P({\mib x})=\frac{1}{3}\tr\{W(x)\}
  \enum
}
(where $W(x)$ denotes the Wilson line that passes through $x$
and wraps around space-time in the time direction) assumes
a non-zero expectation value. The expectation value breaks
the $\gz_3$ center symmetry of the theory and
its phase is spontaneously chosen to be $0$, $2\pi/3$ or
$-2\pi/3$. A technically attractive choice of order parameter
is the Polyakov loop at positive flow time, since
its distribution does not require renormalization
[\ref{RenFlow}] and unambiguously shows the increasingly strong
polarization of the loop with increasing temperature
(see fig.~2).
Like the freezing of the topological charge,
the spontaneous breaking of the center symmetry is associated
with very long autocorrelation times if the standard
simulation algorithms are used.

Master fields representative of the theory in a pure phase
can be built up in several steps from approximately
thermalized configurations on smaller lattices.
If $L$ is not very much larger than $L_0$, the simulation
algorithm rapidly evolves the gauge field to a field with
definite polarization of the Polyakov loop.
Periodic extensions of the field in space to
larger lattices preserve the polarization
and long equilibration times caused by large domains
with different polarization are avoided.
Reflections in space preserve the distribution of the Polyakov loop too and
additionally ensure that the topological charge of the field
and thus its effects on the correlation functions
[\ref{BrowerEtAl},\ref{AokiEtAl}] remain small.

\input figure3

\subsection 3.2 Simulation results

In the continuum limit, the topological susceptibility is independent
of the flow time $t$ at which the charge density $q(x)$ is computed,
provided $t$ is held fixed in physical units when the limit is
taken. The choice of the flow time however has an influence on the
size of the lattice effects. In the calculations reported here,
two values of $t$ given in units of $t_0$ were chosen
corresponding to smoothing ranges $\sqrt{8t}$ [\ref{WilsonFlow}]
approximately equal to $0.28\,\fm$ and $0.47\,\fm$.

As explained in ref.~[\ref{LuscherMaster}], $\chi_t$ can be obtained
in master-field simulations by integrating the two-point
correlation function of the charge density,
\equation{
  \chi_t(R)=a^4\sum_{x_0}\sum_{|{\mib x}|\leq R}\langle q(x)q(0)\rangle,
  \enum
}
up to some sufficiently large radius $R$, where the integral
reaches its asymptotic value within statistical errors
(see fig.~3 for illustration).
Reflection positivity implies that the asymptotic value is approached
from above with an exponential rate given by the screening
lengths in the pseudo-scalar channel.

The bumps in the data shown in fig.~3 and the plateaus at $R\geq1.2\,\fm$
are characteristic features of $\chi_t(R)$ on all lattices
listed in table~2. At large $T$, small $R$ and small flow times $t$,
$\chi_t(R)$ probes the two-point function of the topological density
at short distances, where perturbation theory applies.
The bumps in the data are in fact roughly matched by
leading-order perturbation theory (appendix B).
This computation also shows that $\chi_t(R)$
is suppressed already at small $R$ by the
gradient-flow smoothing of the charge density
and then gets further suppressed at larger radii
by the negative (non-perturbative) long-distance contributions.

The results for the topological susceptibility quoted in table~3
coincide with the calculated values of $\chi_t(R)$ at
$R\simeq1.4\,\fm$, where the asymptotic plateaus are, in all cases,
safely reached within errors.

\input table3

\subsection 3.3 Continuum limit

The calculated values of $t_0^2\chi_t$ must be expected to depend
on the lattice spacing, the leading effects near the continuum limit
being of order $a^2$. Statistically significant
lattice effects are, however, only observed at the larger
temperature considered (see table~3 and fig.~4).
As further elucidated in subsect.~3.4, it is in fact
no suprise that the relative size of the effects increases with temperature,
since the lattice expression for the topological charge density includes
non-topological contributions of order $a^2$.

Linear extrapolation in $a^2/t_0$ of the data listed in table~3
to the continuum limit yield results
for $t_0^2\chi_t$ with errors ranging from $5.3$ to
$14$ percent. The values obtained at the
two flow times considered agree within errors,
as should be the case, the ones at the larger flow time,
\equation{
  t_0^2\chi_t=2.25(12)\times10^{-5}\quad\hbox{at}\quad
  T\sqrt{8t_0}=1.081,
  \enum
  \nexteq{2.5ex}
  t_0^2\chi_t=3.43(27)\times10^{-6}\quad\hbox{at}\quad
  T\sqrt{8t_0}=1.434,
  \enum
}
being a bit more precise.
These figures are orders of magnitude smaller than
the susceptibility $t_0^2\chi_t=6.67(7)\times10^{-4}$ [\ref{CeEtAl}]
at zero temperature and the observed
rapid decrease from $T=1.5\,\Tc$ to $T=2.0\,\Tc$ is in rough
agreement with the power law (2.4).
The agreement might however be somewhat fortuitous
in view of the fact that the derivation of eq.~(2.4) assumes
the effective gauge coupling to be small, which is not the case
at these temperatures.

\input figure4

\subsection 3.4 Miscellaneous remarks

{\it Scaling behaviour}.
If both $T$ and $L$ are held fixed in physical units,
the computational effort required for the generation
of a single master field is expected to increase
like $a^{-6}$ when the continuum limit is approached.
With respect to the integral method,
which scales approximately like $a^{-10}$, this behaviour
is rather mild. However, if $T$ is increased at fixed $a$,
$L$ must grow too for the inequality (1.1) to remain true.
While the computational effort then scales
like $T^7$ or so, the higher cost of the simulations should be balanced
against the fact that the effective statistics provided by
a single master field increases proportionally to $T^8$.

\vskip1ex
\noindent
{\it Improved topological charge}. In all computations reported here,
the standard symmetric expression was used for the topological
charge density on the lattice, in which the field tensor
$F_{\mu\nu}(x)$ is given by the so-called clover formula.
A classically $\rmO(a^2)$-improved expression is then
\equation{
  q(x)=
  -{1\over32\pi^2}\,\epsilon_{\mu\nu\rho\sigma}
  \tr\bigl\{F_{\mu\nu}(x)F_{\rho\sigma}(x)
  -\frac{2}{3}a^2F_{\mu\nu}(x)
  [F_{\mu\rho}(x),F_{\mu\sigma}(x)]
  \bigr\}
  \enum
}
up to derivative terms that do not contribute to the total
charge $Q$. Contrary to what may be expected, the $a^2$-correction
in eq.~(3.5) tends to increase the lattice-spacing dependence of the
topological susceptibility. A complete
$\rmO(a^2)$-improvement of the theory [\ref{OnShell}]
and the gradient flow [\ref{RamosSint}] is thus presumably required
if the convergence to the continuum limit is to be accelerated.

\vskip1ex
\noindent
{\it Finite-volume effects in traditional simulations}.
At high temperatures $T$,
the basic screening lengths are expected to decrease proportionally
to $1/T$. The approximate susceptibility $\chi_t(R)$
therefore approaches its asymptotic value at large $R$ more
and more rapidly, but as suggested by fig.~3, a significant $R$-dependence
may persist in a core range of $R$ extending up to $R=1.2\,\fm$ or so.
In traditional high-temperature simulations,
where the topology freezing is overcome in ways other than
through a large volume, spatial sizes
$L\geq2.4\,\fm$ are thus required to be safe of finite-volume effects.

%% file: table2
\topinsert
\newdimen\digitwidth
\setbox0=\hbox{\rm 0}
\digitwidth=\wd0
\catcode`@=\active
\def@{\kern\digitwidth}
\tablecaption{Parameters of the high-temperature lattices}
\vskip-3.5ex

$$\vbox{\settabs\+&%
                  xxxxxxx&xx&%
                  xxxxxxxxx&xx&%
                  xxxxxxxxx&xx&%
                  xxxx&xx&%
                  xxxxxxxx&xx&%
                  xxxxxxxx&xx&%
                  xxxxxxxx&x\cr
\thicktablerule
\vskip1.2ex
                \+& \hfill Label\hfill
                 && \hfill Lattice\hfill
                 && \hfill $\beta$\hfill
                 && \hfill $N_{\rm mf}$\hfill
                 && \hfill $a\,[\fm]$\hfill
                 && \hfill $T\,[\MeV]$\hfill
                 && \hfill $L\,[\fm]$\hfill
                 &\cr
\vskip0.8ex
\thintablerule
\vskip1.2ex
{\ninepoint
  \+& \hfill ${\rm A}_1$\hfill
  &&  \hfill $@6\times256^3$\hfill
  &&  \hfill $6.15533$\hfill
  &&  \hfill $10$\hfill
  &&  \hfill $0.073$\hfill
  &&  \hfill $449.1$\hfill
  &&  \hfill $18.7$\hfill
  &\cr
  \+& \hfill ${\rm A}_2$\hfill
  &&  \hfill $@8\times384^3$\hfill
  &&  \hfill $6.35393$\hfill
  &&  \hfill $10$\hfill
  &&  \hfill $0.055$\hfill
  &&  \hfill $449.1$\hfill
  &&  \hfill $21.1$\hfill
  &\cr
  \+& \hfill ${\rm A}_3$\hfill
  &&  \hfill $12\times512^3$\hfill
  &&  \hfill $6.65454$\hfill
  &&  \hfill $30$\hfill
  &&  \hfill $0.037$\hfill
  &&  \hfill $449.1$\hfill
  &&  \hfill $18.7$\hfill
  &\cr
\vskip0.5ex
  \+& \hfill ${\rm B}_1$\hfill
  &&  \hfill $@6\times512^3$\hfill
  &&  \hfill $6.35033$\hfill
  &&  \hfill $18$\hfill
  &&  \hfill $0.055$\hfill
  &&  \hfill $595.8$\hfill
  &&  \hfill $28.3$\hfill
  &\cr
  \+& \hfill ${\rm B}_2$\hfill
  &&  \hfill $@8\times768^3$\hfill
  &&  \hfill $6.56185$\hfill
  &&  \hfill $20$\hfill
  &&  \hfill $0.041$\hfill
  &&  \hfill $595.8$\hfill
  &&  \hfill $31.8$\hfill
  &\cr
  \+& \hfill ${\rm B}_3$\hfill
  &&  \hfill $@12\times1024^3$\hfill
  &&  \hfill $6.87251$\hfill
  &&  \hfill $20$\hfill
  &&  \hfill $0.028$\hfill
  &&  \hfill $595.8$\hfill
  &&  \hfill $28.3$\hfill
  &\cr
}
\vskip1.0ex
\thicktablerule
}
$$
\vskip-1.0ex
\endinsert

%% file: figure2
\topinsert
\vbox{
\vskip0ex

\epsfxsize=11.0cm\hskip0.5cm\epsfbox{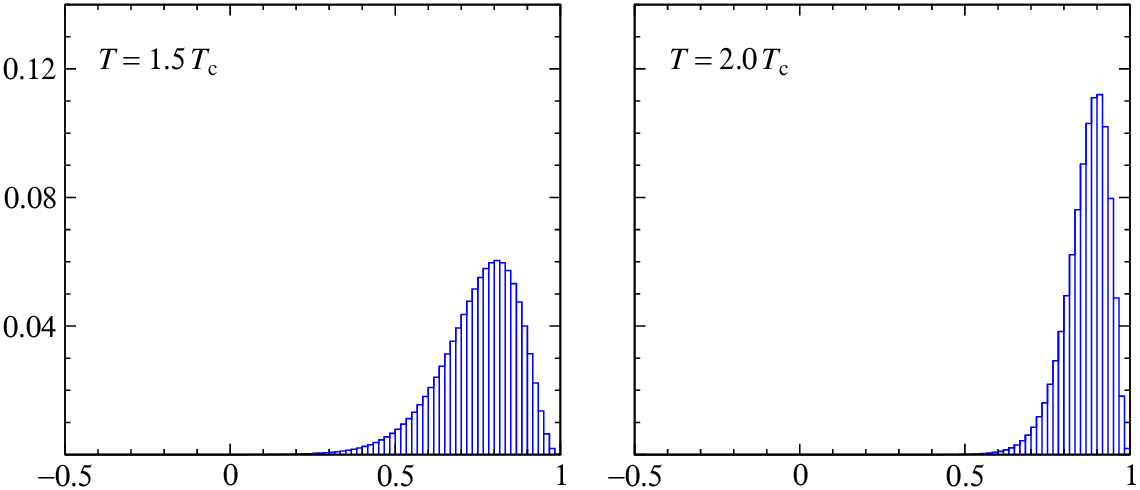}

\vskip2.0ex
\figurecaption{%
Normalized histograms of $\Re\{zP(\xv)\}$ at
flow time $t=0.35\,t_0$ measured on the
$A_1$ (left) and $B_1$ (right) lattices. In both cases,
the bin size is $1/60$ and
the phase factor $z\in\{1,\exp(\pm i2\pi/3)\}$ is chosen so as
to cancel the phase of the average value of the Polyakov loop.
}
}
\vskip2ex
\endinsert

%% file: figure3
\topinsert
\vbox{
\vskip0ex

\epsfxsize=8.0cm\hskip2.0cm\epsfbox{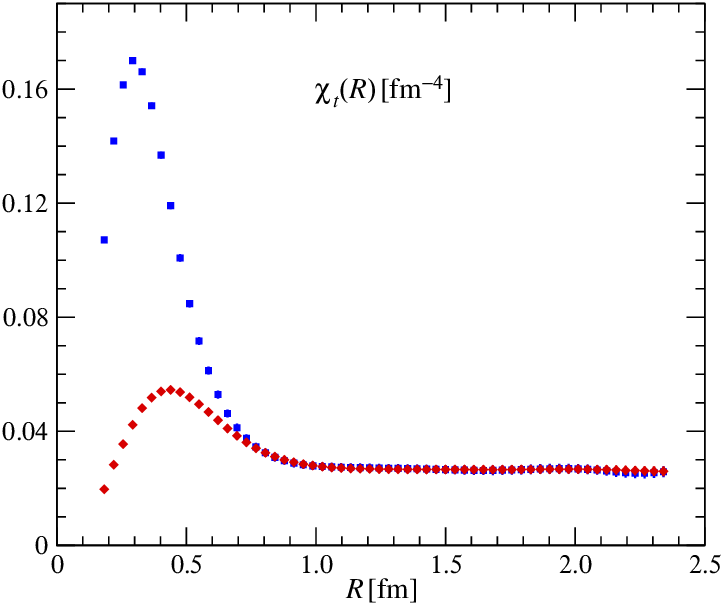}

\vskip2.0ex
\figurecaption{%
Values of $\chi_t(R)$ obtained on the $A_3$ lattice
at two flow times corresponding to smoothing ranges equal to
$0.28\,\fm$ (squares) and $0.47\,\fm$ (diamonds).
}
}
\vskip-1ex
\endinsert

%% file: table3
\topinsert
\newdimen\digitwidth
\setbox0=\hbox{\rm 0}
\digitwidth=\wd0
\catcode`@=\active
\def@{\kern\digitwidth}
\tablecaption{Simulation results for $\chi_t$}
\vskip-3.5ex

$$\vbox{\settabs\+&%
                  xxxxx&xx&%
                  xxxxx&xx&%
                  xxxxxxxx&x&%
                  xxxxxxxxxxxx&xxx&%
                  xxxxxxxx&x&%
                  xxxxxxxxxxx&\cr
\thicktablerule
\vskip1.2ex
                \+& \hfill Run\hfill
                 && \hfill $R/a$\hfill
                 && \hfill $\sqrt{t/t_0}$\hfill
                 && \hfill $t_0^2\chi_t\times10^5$\hfill
                 && \hfill $\sqrt{t/t_0}$\hfill
                 && \hfill $t_0^2\chi_t\times10^5$\hfill
                 &\cr
\vskip0.8ex
\thintablerule
\vskip1.2ex
{\ninepoint
  \+& \hfill ${\rm A}_1$\hfill
  &&  \hfill $20$\hfill
  &&  \hfill $0.590$\hfill
  &&  \hfill $2.233(89)$\hfill
  &&  \hfill $0.983$\hfill
  &&  \hfill $2.089(75)$\hfill
  &\cr
  \+& \hfill ${\rm A}_2$\hfill
  &&  \hfill $26$\hfill
  &&  \hfill $0.590$\hfill
  &&  \hfill $2.33(10)@$\hfill
  &&  \hfill $0.983$\hfill
  &&  \hfill $2.281(79)$\hfill
  &\cr
  \+& \hfill ${\rm A}_3$\hfill
  &&  \hfill $39$\hfill
  &&  \hfill $0.590$\hfill
  &&  \hfill $2.12(12)@$\hfill
  &&  \hfill $0.983$\hfill
  &&  \hfill $2.11(11)@$\hfill
  &\cr
\vskip0.5ex
  \+& \hfill ${\rm B}_1$\hfill
  &&  \hfill $26$\hfill
  &&  \hfill $0.593$\hfill
  &&  \hfill $0.494(26)$\hfill
  &&  \hfill $0.988$\hfill
  &&  \hfill $0.402(14)$\hfill
  &\cr
  \+& \hfill ${\rm B}_2$\hfill
  &&  \hfill $34$\hfill
  &&  \hfill $0.593$\hfill
  &&  \hfill $0.400(20)$\hfill
  &&  \hfill $0.988$\hfill
  &&  \hfill $0.372(12)$\hfill
  &\cr
  \+& \hfill ${\rm B}_3$\hfill
  &&  \hfill $52$\hfill
  &&  \hfill $0.593$\hfill
  &&  \hfill $0.343(38)$\hfill
  &&  \hfill $0.988$\hfill
  &&  \hfill $0.370(32)$\hfill
  &\cr
}
\vskip1.0ex
\thicktablerule
}
$$
\vskip-1.0ex
\endinsert

%% file: figure4
\topinsert
\vbox{
\vskip0ex

\epsfxsize=11.0cm\hskip0.5cm\epsfbox{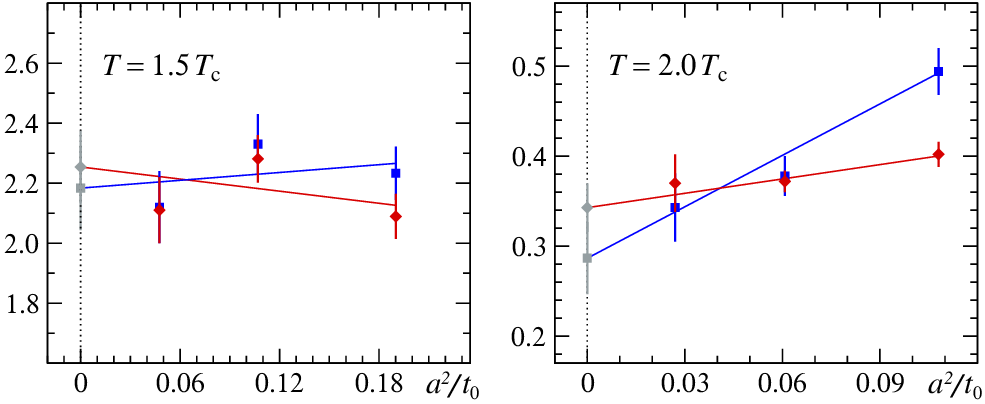}

\vskip3.0ex
\figurecaption{%
Extrapolation of the values of $t_0^2\chi_t\times10^5$
listed in table~3 to the
continuum limit (left: $A$-lattices, right: $B$-lattices).
The data at flow time $0.35\,t_0$ (squares) and
$0.97\,t_0$ (diamonds) are extrapolated linearly in $a^2$,
the grey points at $a=0$ being the extrapolated values.
}
}
\endinsert

%% file: sect4
\section 4. Conclusions

Dimensional analysis suggests that
the topological susceptibility grows proportionally to $T^4$
at high temperatures $T$, but instead it decreases rapidly
as a result of a nearly perfect cancellation of
short- and long-distance contributions.
This behaviour is commonly attributed to the topological nature
of the charge density $q(x)$, i.e.~to the fact that variations
of $q(x)$ with respect to the gauge field are total derivatives.
None of the non-perturbatively well-defined expressions
for the susceptibility known to date however embodies this property of
the charge density to the extent that the smallness of the susceptibility
at high temperatures would be explained.

Master-field simulations provide new opportunities for non-perturbative
studies of QCD. At non-zero temperatures below $\Tc$, for example,
the physically large volumes that become accessible in this way
allow the theory to be studied in kinematic regimes close to
the thermodynamic limit, where multi-hadron states make
important contributions to the partition function.
Another motivation for the use of this new type of
simulations is the fact that the
topology-freezing issue (which tends to become severe at
lattice spacings $a\leq0.05\,\fm$)
can be bypassed in a conceptually transparent manner.

The computations of the topological susceptibility
reported in the present paper could proceed straightforwardly
for this reason and led to results with unprecedented precision.
At temperatures higher than the ones considered here,
master-field simulations however require
larger and larger lattices to be simulated
and thus become impractical at some point.
Moreover, the topological susceptibility
must be expected to be increasingly sensitive to
lattice effects. To be able to control these effects,
the lattice spacing must then be decreased.
This second problem is, however, not specific to master-field simulations
and will persist until an expression for the susceptibility
is found which is naturally small at high temperatures.

\vskip1ex
All simulations were performed on a HPC cluster at CERN and on
the Marconi machine at CINECA through agreements of INFN
and the University of Milano-Bicocca with CINECA.
We gratefully acknowledge the computer resources and
the technical support provided by these institutions.

%


%% file: appa
\appendix A. Simulation algorithm and other implementation details

Apart from some specific technical details related to
the very large sizes of the simulated lattices,
the master-field simulations reported in this paper
followed established lattice-QCD strategies.

\subsection A.1 Simulation algorithm

All simulations were performed using the HMC algorithm [\ref{HMC}] with
trajectory length $\tau=2$. The
molecular-dynamics equations were integrated by applying
the forth-order integrator given by eqs.~(63) and (71) in ref.~[\ref{OMF}].
This scheme proves to be highly efficient and an only mild
adjustment of the step number $n_{\rm step}$ was required on the larger
lattices in order to preserve a good acceptance rate $P_{\rm acc}$ (see table~4).

Using standard MPI communication functions,
the computational work was distributed over
up to $32768$ processing units.
Most demanding from the point of view of the memory requirements was
the measurement program for the topological susceptibility,
which occupied a total memory of about $16$ TB in the case of the largest
lattice.

\input table4

\subsection A.2 Thermalization

As already indicated in sect.~3, the master fields
were generated in several steps from smaller lattices, where
thermalizations of the gauge field alternate with
extensions to the next larger lattice
through reflections at the lattice planes.
The plaquette action per point is unchanged
after a reflection and the topological charge vanishes, but
the gauge-field tensor changes abruptly
across the reflection planes, which can give rise to
a low acceptance rate in the early phase of the
subsequent thermalization.
A few update cycles with a more
accurate integration of the molecular-dynamics equations
may be required in this case to get the thermalization
started.

The lengths $\tau_{\rm th}$ of the final thermalization runs
listed in table~4 are much longer than the relevant
autocorrelation times.
A drift in the single-field expectation values [\ref{LuscherMaster}]
has in fact never been seen after these long thermalization phases
(see fig.~5 for an example). It may be
worth noting in passing that outliers, such as the
measurement number 25 in fig.~5,
must occur with some non-zero probability, as in traditional simulations,
where whole ensemble averages may be similarly outlying.

The separation $\Delta\tau_{\rm mf}$
in simulation time of the master fields included in the
computations of expectation values need not be
particularly large, since any statistical correlations among
the fields are automatically taken into account [\ref{LuscherMaster}].
Autocorrelations however lead to larger statistical errors relative to
what they would be for uncorrelated fields.
On the $B_3$ lattice, for example,
the separation was duplicated with respect to the
other runs for this reason.

\input figure5

\subsection A.3 Use of quadruple-precision arithmetic

On the simulated lattices,
significance losses of up to 11 decimal places
occur when the energy deficit $\Delta H$
is computed at the end of the molecular-dynamics evolution
of the fields.
Standard IEEE 754 double-precision data and arithmetic may be barely
good enough under these conditions and it is, therefore, advisable
to use quadruple-precision artithmetic in the summation of the
action densities over all lattice points.
$\Delta H$ is then obtained with absolute precision
given by the now practically exactly
accumulated numerical errors of the densities.
Assuming these are randomly distributed, their
sum scales like $(V/a^4)^{1/2}$ and the accumulated
inaccuracies are then far below any statistically relevant level.

A convenient portable implementation of quadruple-precision
numbers is through pairs of double-precision numbers. Algorithms
for the associated arithmetic operations were
published by Dekker [\ref{Dekker}] many years ago. The subject
is also discussed in a book of Knuth [\ref{Knuth}] and
more extensively in an article by Shewchuk [\ref{Shewchuk}].

\subsection A.4 Parallel\/ {\rm I/O}

In master-field simulations, the computer time spent for field
configuration I/O may not be negligible.
Current HPC systems however permit the storage facilities
to be accessed concurrently and thus offer a high aggregate
I/O bandwidth.

In the I/O programs used in the present study, the lattice is
logically divided into fairly large rectangular blocks. The
part of the gauge field
residing on a given block is then written out in a portable format by
one of the processing units. A single field is thus stored in
several files and advantage of the parallel capabilities of the
storage facility is taken by having many processing units write
their blocks concurrently.

%% file: table4
\topinsert
\newdimen\digitwidth
\setbox0=\hbox{\rm 0}
\digitwidth=\wd0
\catcode`@=\active
\def@{\kern\digitwidth}
\tablecaption{Simulation parameters$\,^{\ast}$}
\vskip-1.0ex

$$\vbox{\settabs\+&%
                  xxxxxxxx&xx&%
                  xxxxxxxx&xx&%
                  xxxxxxxx&xx&%
                  xxxxxxxx&xx&%
                  xxxxxxxx&\cr
\thicktablerule
\vskip1.2ex
                \+& \hfill Run\hfill
                 && \hfill $n_{\rm step}$\hfill
                 && \hfill $P_{\rm acc}$\hfill
                 && \hfill $\tau_{\rm th}$\hfill
                 && \hfill $\Delta\tau_{\rm mf}$\hfill
                 &\cr
\vskip0.8ex
\thintablerule
\vskip1.2ex
{\ninepoint
  \+& \hfill ${\rm A}_1$\hfill
  &&  \hfill $13$\hfill
  &&  \hfill $0.95$\hfill
  &&  \hfill $15360$\hfill
  &&  \hfill $480$\hfill
  &\cr
  \+& \hfill ${\rm A}_2$\hfill
  &&  \hfill $13$\hfill
  &&  \hfill $0.91$\hfill
  &&  \hfill $10560$\hfill
  &&  \hfill $480$\hfill
  &\cr
  \+& \hfill ${\rm A}_3$\hfill
  &&  \hfill $17$\hfill
  &&  \hfill $0.93$\hfill
  &&  \hfill $@8160$\hfill
  &&  \hfill $480$\hfill
  &\cr
\vskip0.5ex
  \+& \hfill ${\rm B}_1$\hfill
  &&  \hfill $13$\hfill
  &&  \hfill $0.87$\hfill
  &&  \hfill $@3840$\hfill
  &&  \hfill $480$\hfill
  &\cr
  \+& \hfill ${\rm B}_2$\hfill
  &&  \hfill $17$\hfill
  &&  \hfill $0.89$\hfill
  &&  \hfill $@6240$\hfill
  &&  \hfill $480$\hfill
  &\cr
  \+& \hfill ${\rm B}_3$\hfill
  &&  \hfill $18$\hfill
  &&  \hfill $0.85$\hfill
  &&  \hfill $@6080$\hfill
  &&  \hfill $960$\hfill
  &\cr
}
\vskip1.0ex
\thicktablerule
\vskip1.0ex
\+&{\footnotefont$^{\ast}\;\tau_{\rm th}$ and $\Delta\tau_{\rm mf}$ are given in
units of molecular-dynamics time}\cr
}
$$
\endinsert

%% file: figure5
\topinsert
\vbox{
\vskip0ex

\epsfxsize=8.0cm\hskip2.0cm\epsfbox{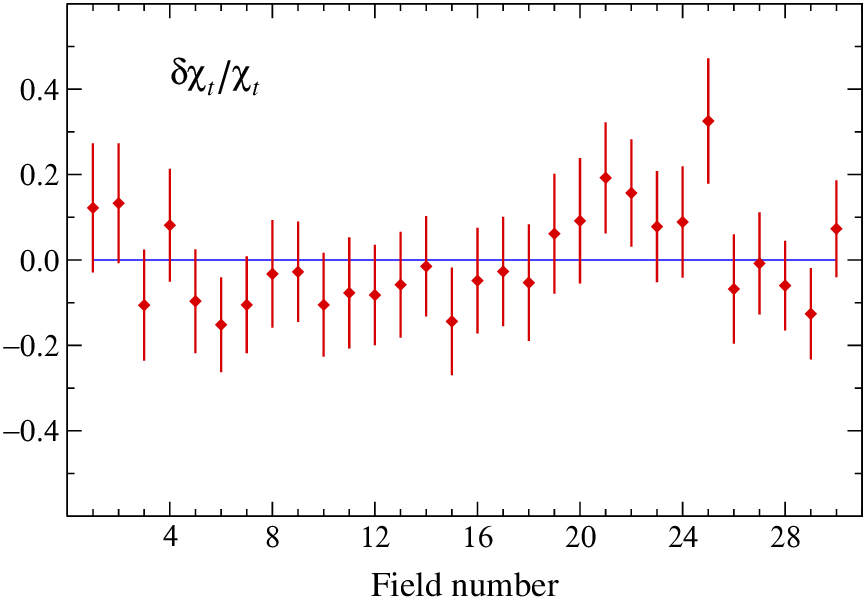}

\vskip3.0ex
\figurecaption{%
Relative deviation from the ensemble average
of the values of the topological susceptibility
computed using single master fields
(run $A_3$, flow time $0.97\,t_0$).
}
}
\endinsert

%% file: appb
\appendix B. Calculation of $\chi_t(R)$ in perturbation theory

In the continuum theory and at flow time $t>0$,
the integrated correlation function
\equation{
  \chi_t(R)=\int_0^{1/T}\rmd x_0\int_{|{\mib x}|\leq R}\rmd^3{\mib x}\,
  \left\langle q(x)q(0)\right\rangle
  \enum
}
of the topological charge density
can be straightforwardly expanded in powers of the gauge coupling.
The computation proceeds along the lines of
ref.~[\ref{WilsonFlow}] except for the fact that
the time components $p_0$ of the momenta are quantized in units
of $2\pi T$.

At high temperatures, where
\equation{
  8tT^2\pi^2\gg1,
  \enum
}
the contributions of the $p_0\neq0$ modes of the gauge field
to the leading-order expression for the two-point function of
the charge density are exponentially suppressed. Up to
these terms the latter is then given by
\equation{
  \left\langle q(x)q(0)\right\rangle=
  \alpha_s^2{T^2\over\pi^5(8t)^3r^3}\gamma(\frac{3}{2},r)
  \bigl\{3\gamma(\frac{3}{2},r)
  -4\gamma(\frac{5}{2},r)\bigr\},
  \enum
}
$\alpha_s$ being the strong coupling,
\equation{
  \gamma(a,r)=\int_0^r\rmd s\,s^{a-1}\rme^{-s}
  \enum
}
the incomplete $\Gamma$-function and $r={\mib x}^2/(8t)$.
The correlation function thus decreases
monotonically from
\equation{
  \left\langle q(0)^2\right\rangle=\alpha_s^2{4T^2\over3\pi^5(8t)^3},
  \enum
}
becomes negative at some point and eventually goes to zero with a rate
proportional to $|{\mib x}|^{-6}$ at large distances $|{\mib x}|$.

\input figure6

Equation (B.3) leads to the expression
\equation{
  \chi_t(R)=\alpha_s^2{4T\over\pi^4(8t)^{3/2}}f(\rho),
  \qquad
  \rho={R\over(8t)^{1/2}},
  \enum
  \nexteq{2.5ex}
  f(\rho)=\rho^{-3}\gamma(\frac{3}{2},\rho^2)^2,
  \enum
}
for the approximate susceptibility (B.1). To this order of perturbation
theory, $\chi_t(R)$ thus depends on the summation radius $R$ roughly
like the data plotted in fig.~3 (see fig.~6).
In particular,
at the flow times chosen in the simulations, the maxima
of the bumps in fig.~3 are at $R=0.30\,\fm$ and $0.45\,\fm$,
while the leading-order expression (B.6) has its maximum
at $R=0.35\,\fm$ and $0.58\,\fm$ in these cases.
The plateaus in fig.~3, on other hand, occur at distances,
where perturbation theory is not expected to apply and
instead goes to zero consistently with the vanishing of $\chi_t$
to all orders.

%% file: figure6
\topinsert
\vbox{
\vskip0ex

\epsfxsize=6.0cm\hskip3.0cm\epsfbox{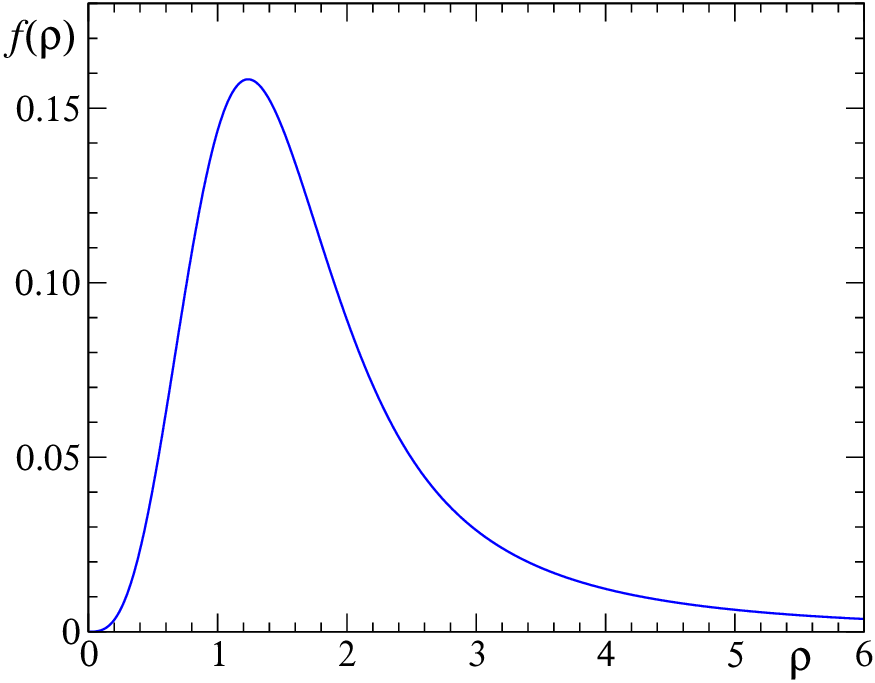}

\vskip3.0ex
\figurecaption{%
The function (B.7)
assumes its maximal value $0.1583(1)$ at $\rho=1.236(1)$
and decays like $\rho^{-3}$ at large $\rho$.
}
}
\endinsert

%% file: biblio
\beginbibliography


\bibitem{PecceiQuinnI}
R. D. Peccei, H. R. Quinn,
{\it CP conservation in the presence of instantons},
Phys. Rev. Lett. 38 (1977) 1440


\bibitem{PecceiQuinnII}
R. D. Peccei, H. R. Quinn,
{\it Constraints imposed by CP conservation in the presence of instantons},
Phys. Rev. D16 (1977) 1791


\bibitem{Weinberg}
S. Weinberg,
{\it A new light boson?},
Phys. Rev. Lett. 40 (1978) 223


\bibitem{Wilczek}
F. Wilczek,
{\it Problem of strong $P$ and $T$ invariance in the presence of
instantons},
Phys. Rev. Lett. 40 (1978) 279



\bibitem{BerkowitzEtAl}
E. Berkowitz, M. I. Buchoff, E. Rinaldi,
{\it Lattice QCD input for axion cosmology},
Phys. Rev. D 92 (2015) 034507


\bibitem{BorsanyiEtAlI}
S. Borsanyi et al.,
{\it Axion cosmology, lattice QCD and the dilute instanton gas},
Phys. Lett. B752 (2016) 175


\bibitem{BonatiEtAlI}
C. Bonati et al.,
{\it Axion phenomenology and $\theta$-dependence from $N_f=2+1$
lattice QCD},
JHEP 1603 (2016) 155


\bibitem{PetreczkyEtAl}
P. Petreczky, H. P. Schadler, S. Sharma,
{\it The topological susceptibility in finite temperature QCD and
axion cosmology},
Phys. Lett. B762 (2016) 498


\bibitem{BorsanyiEtAlII}
 S. Borsanyi et al.,
{\it Calculation of the axion mass based on high-temperature lattice
quantum chromodynamics},
Nature 539 (2016) 69


\bibitem{FrisonEtAl}
J. Frison, R. Kitano, H. Matsufuru, S. Mori, N. Yamada,
{\it Topological susceptibility at high temperature on the lattice},
JHEP 1609 (2016) 021

\bibitem{TaniguchiEtAl}
Y. Taniguchi, K. Kanaya, H. Suzuki, T. Umeda,
{\it Topological susceptibility in finite temperature (2+1)-flavor QCD
using gradient flow},
Phys. Rev. D95 (2017) 054502

\bibitem{JahnEtAl}
P. T. Jahn, G. D. Moore, D. Robaina,
{\it $\chi_{\rm top}(T\gg T_{\rm c})$ in pure-glue QCD through reweighting},
Phys. Rev. D98 (2018) 054512

\bibitem{BurgerEtAl}
F. Burger, E. M. Ilgenfritz, M. P. Lombardo, A. Trunin,
{\it Chiral observables and topology in hot QCD with two families of quarks},
Phys. Rev. D98 (2018) 094501

\bibitem{BonatiEtAlII}
C. Bonati et al.,
{\it Topology in full QCD at high temperature: a multicanonical approach},
JHEP 1811 (2018) 170


\bibitem{LuscherMaster}
M. L\"uscher,
{\it Stochastic locality and master-field simulations of very large lattices},
EPJ Web Conf. 175 (2018) 01002



\bibitem{BrowerEtAl}
R. Brower, S. Chandrasekharan, J.W. Negele, U.-J. Wiese,
{\it QCD at fixed topology},
Phys. Lett. B560 (2003) 64

\bibitem{AokiEtAl}
S. Aoki, H. Fukaya, S. Hashimoto, T. Onogi,
{\it Finite volume QCD at fixed topological charge},
Phys. Rev. D76 (2007) 054508



\bibitem{WilsonFlow}
M. L\"uscher,
{\it Properties and uses of the Wilson flow in lattice QCD},
JHEP 1008 (2010) 071 [Erratum: {\it ibid.} 1403 (2014) 092]

\bibitem{RenFlow}
M. L\"uscher, P. Weisz,
{\it Perturbative analysis of the gradient flow in non-Abelian gauge
theories},
JHEP 1102 (2011) 051


\bibitem{Wilson}
K. G. Wilson, {\it Confinement of quarks}, Phys. Rev. D10 (1974) 2445


\bibitem{GiustiEtAlI}
L. Giusti, G. C. Rossi, M. Testa, G. Veneziano,
{\it The $U_{\rm A}(1)$ problem on the lattice with Ginsparg--Wilson fermions},
Nucl. Phys. B628 (2002) 234


\bibitem{GiustiEtAlII}
L. Giusti, G. C. Rossi, M. Testa,
{\it Topological susceptibility in full QCD with Ginsparg--Wilson fermions},
Phys. Lett. B587 (2004) 157


\bibitem{LuscherShort}
M. L\"uscher,
{\it Topological effects in QCD and the problem of short-distance
singularities},
Phys. Lett. B593 (2004) 296


\bibitem{CeEtAl}
M. C\`e, C. Consonni, G. P. Engel, L. Giusti,
{\it Non-Gaussianities in the topological charge distribution of
the SU(3) Yang--Mills theory},
Phys. Rev. D92 (2015) 074502



\bibitem{SemiClassical}
M. L\"uscher,
{\it A semiclassical formula for the topological susceptibility
in a finite space-time volume},
Nucl. Phys. B205 [FS5] (1982) 483

\bibitem{BaalKraanI}
T. C. Kraan, P. van Baal,
{\it Periodic instantons with nontrivial holonomy},
Nucl. Phys. B533 (1998) 627

\bibitem{BaalKraanII}
T. C. Kraan, P. van Baal,
{\it Monopole constituents inside SU(n) calorons},
Phys. Lett. B435 (1998) 389


\bibitem{SommerScale}
R. Sommer,
{\it A new way to set the energy scale in lattice gauge theories
and its applications to the static force and $\alpha_s$ in
SU(2) Yang--Mills theory},
Nucl. Phys. B411 (1994) 839

\bibitem{GuagnelliEtAl}
M. Guagnelli, R. Sommer, H. Wittig (ALPHA collab.),
{\it Precision computation of a low-energy reference scale in
quenched lattice QCD},
Nucl. Phys. B535 (1998) 389

\bibitem{NeccoSommer}
S. Necco, R. Sommer,
{\it The $N_f=0$ heavy quark potential from short to intermediate distances},
Nucl.Phys. B622 (2002) 328


\bibitem{AsakawaEtAl}
M. Asakawa et al.,
{\it Determination of reference scales for Wilson gauge action
from Yang--Mills gradient flow},
arXiv:1503.06516


\bibitem{BoydEtAl}
G. Boyd et al.,
{\it Thermodynamics of SU(3) lattice gauge theory},
Nucl. Phys. B469 (1996) 419


\bibitem{OnShell}
M. L\"uscher, P. Weisz
{\it On-shell improved lattice gauge theories},
Commun. Math. Phys. 97 (1985) 59 [E: {\it ibid.} 98 (1985) 433]

\bibitem{RamosSint}
A. Ramos, S. Sint
{\it Symanzik improvement of the gradient flow in lattice gauge theories},
Eur. Phys. J. C76 (2016) 15



\bibitem{HMC}
S. Duane, A. D. Kennedy, B. J. Pendleton, D. Roweth,
{\it Hybrid Monte Carlo},
Phys. Lett. B195 (1987) 216


\bibitem{OMF}
I. P. Omelyan, I. M. Mryglod, R. Folk,
{\it Symplectic analytically integrable decomposition
algorithms: classification, derivation, and application to molecular
dynamics, quantum and celestial mechanics simulations},
Comp. Phys. Commun. 151 (2003) 272




\bibitem{Dekker}
T. J. Dekker,
{\it A floating-point technique for extending the available precision},
Numer. Math. 18 (1971) 224

\bibitem{Knuth}
D. E. Knuth,
{\it Semi-Numerical Algorithms}, in:
The Art of Computer Programming, vol. 2, 2nd ed.
(Addison-Wesley, Reading MA, 1981)

\bibitem{Shewchuk}
J. R. Shewchuk, {\it Adaptive precision floating-point arithmetic
and fast robust geo\-metric predicates},
Discrete \& Computational Geometry 18 (1997) 305

\endbibliography